\documentstyle[12pt,epsf,epsfig,wrapfig]{article}
\newcommand{\ba}{\begin{eqnarray}}\newcommand{\ea}{\end{eqnarray}}
\newcommand{\nn}{\nonumber}
\newcommand{\be}{\begin{equation}}
\newcommand{\ee}{\end{equation}}

\def\xb{\overline{x}}

\def\veps{\varepsilon}
\def\als{\alpha_s}
\def\vk{{\bf k}_{\perp}}

\def\gev{\,{\rm GeV}}
\begin{document}

\begin{center}
{\bfseries SPIN EFFECTS AND AMPLITUDE STRUCTURE IN VECTOR MESON
PHOTOPRODUCTION AT SMALL $x$}

\vskip 5mm S.V.Goloskokov $^{\dag}$

\vskip 5mm {\small {\it Bogoliubov Laboratory of Theoretical
Physics, Joint Institute
for Nuclear Research,\\ Dubna 141980, Moscow region, Russia}}\\
$\dag$ {\it E-mail:goloskkv@theor.jinr.ru }
\end{center}

\vskip 5mm
\begin{abstract}
An analysis of light vector meson photoproduction at small $x$ on
the basis of the generalized parton distribution (GPD) approach is
presented. Our results on the cross section and spin density
matrix elements (SDME) are in fair agreement with DESY
experiments. The predicted double spin longitudinal $A_{LL}$
asymmetry  is not small at HERMES energies.
\end{abstract}

\vskip 8mm

This report is devoted to the study of the vector meson
leptoproduction at large energies and small $x$-Bjorken ($x$)
based on our  results \cite{gk05}. In this kinematic region the
process factorizes \cite{fact} into a hard meson leptoproduction
off gluons and GPD (Fig.1).

\begin{wrapfigure}[14]{hr}{5.5cm}
\begin{center}
\mbox{\epsfig{figure=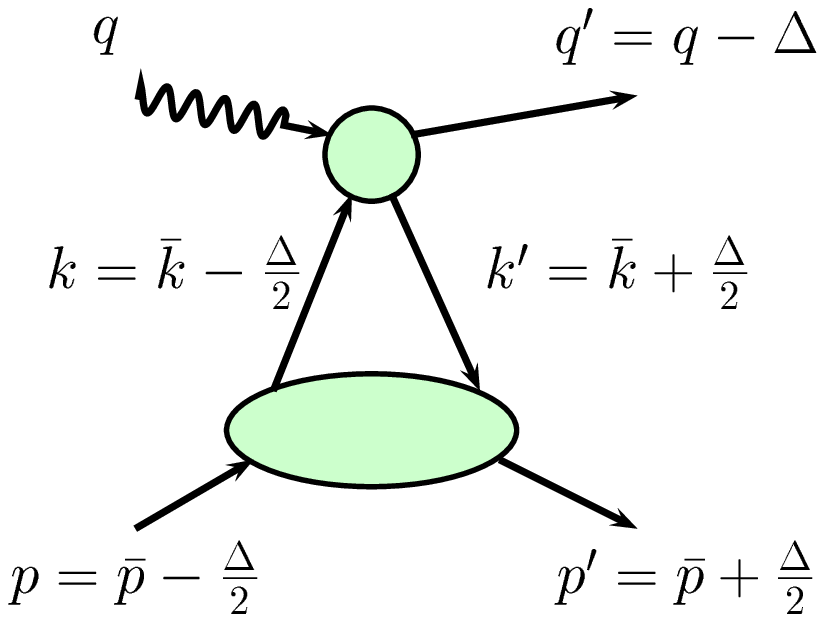,width=5.2cm,height=4cm}}
\end{center}
{\small{\bf Figure 1.} The handbag diagram for the meson
electroproduction off proton.} \label{fig1}
\end{wrapfigure}
The cross section for longitudinally polarized virtual photons
which  dominate for asymptotically large photon virtualities
exceeds the data by a large factor at small $x$ if calculated in
the collinear approximation \cite{mpw}. Moreover, in this
approximation the amplitudes for transversally polarized photons
exhibit infrared singularities \cite{mp} which signal the
breakdown of factorization. Nevertheless the knowledge of  these
amplitudes is necessary to study spin effects in the vector meson
production. Really, the analysis of  SDME for  light vector
mesons, measured by H1 \cite{h1} and ZEUS \cite{zeus}, reveals
that both the $\gamma _{\perp }^{\ast }\rightarrow V_{\perp }$
(TT) and  $\gamma _{\perp }^{\ast }\rightarrow V_{L}$ (TL)
transition amplitudes are non-negligible in the kinematic region
accessible by the DESY experiments. It has been shown by us
\cite{gk04} that in the modified perturbative approach (MPA)
\cite{sterman} which includes the quark transverse degrees of
freedom accompanied by Sudakov suppressions, one can solve both
these problems. Reasonable agreement with H1 and ZEUS data
\cite{h1,zeus} has been obtained for electroproduced  $\rho$ and
$\phi$  mesons at small $x$ \cite{gk05, gk04}.

In this report, we  discuss the spin effects in the vector meson
leptoproduction. Within the MPA the amplitudes for the three
transitions $LL$, $TT$ and $TL$ can be calculated and afterwards
cross sections and SDME. Note that the leading-twist wave function
describes the longitudinally polarized vector mesons and cannot be
applied to the transversally polarized light mesons. To study this
case, we use the higher order $k$- dependent wave function
proposed in \cite{koerner}

The gluon contribution to the leptoproduction amplitudes for $t
\sim 0$ and positive proton helicity reads as a convolution of the
hard subprocess amplitude  ${\cal H}^V$ and a large distance gluon
GPD  $ H^g$ (Fig.1) :
 \ba
{\cal M}_{\mu'+,\mu +} &=& \frac{e}{2}\, {\cal C}_V\,
         \int_0^1 \frac{d\xb}
        {(\xb+\xi)
                  (\xb-\xi + i{\veps})}\nn\\
        &\times& \left\{\, \left[\,{\cal H}^V_{\mu'+,\mu +}\,
         + {\cal H}^V_{\mu'-,\mu -}\,\right]\,
                                   H^g(\xb,\xi,t) \right. \nn\\
       &+& \hspace{2.5mm}\left. \left[\,{\cal H}^V_{\mu'+,\mu +}\,
            -   {\cal H}^V_{\mu'-,\mu -}\,\right]\,
                        \widetilde{H}^g(\xb,\xi,t)\, \right\}\,.
\label{amptt-nf-ji} \ea Here $\mu$ ($\mu'$) denotes the helicity
of the photon (meson), $\xb$ is the  momentum fraction of the
transversally polarized gluons and the skewness $\xi$ is related
to Bjorken-$x$ by $\xi\simeq x/2$. The flavor factors are
$C_{\rho}=1/\sqrt{2}$ and ${ C}_{\phi}=-1/3$.  The polarized GPD
$\widetilde{H}^g$ is much smaller than the GPD $H^g$ at small
$\xb$ and will be important in the $A_{LL}$ asymmetry only.

 The subprocess amplitude ${\cal H}^V$ is represented as  the contraction of the hard
  part $F$, which is calculated perturbatively, and the
non-perturbative meson  wave function $ \phi_V$
\be {\cal
H}^V_{\mu'+,\mu +}\,\pm\,  {\cal H}^V_{\mu'-,\mu -}\,=
\,\frac{2\pi \als(\mu_R)}
           {\sqrt{2N_c}} \,\int_0^1 d\tau\,\int \frac{d^{\,2} \vk}{16\pi^3}
            \phi_{V\mu^\prime}(\tau,k^2_\perp)\;
                  F_{\mu^\prime\mu}^\pm .
\label{hsaml}\ee  The wave function is chosen  in Gaussian form

 \be
          \phi_V(\vk,\tau)\,=\, 8\pi^2\sqrt{2N_c}\, f_V a^2_V
       \, \exp{\left[-a^2_V\, \frac{\vk^{\,2}}{\tau\bar{\tau}}\right]}\,.
\label{wave-l} \ee Here $\tau$ ($\bar{\tau}=1-\tau$) is the
fraction of the meson momentum carried by the quark (antiquark),
$f_V$ is the decay coupling constant and the $a_V$ parameter
determines the value of average transverse momentum of the quark
in the vector meson. Generally, values of $f_V, a_V$ may be
different for the TT and LL amplitudes.

The subprocess amplitude is calculated within the MPA
\cite{sterman} where we keep the $k^2_\perp$ terms in denominators
of the amplitudes and numerator of the TT amplitude. The gluonic
corrections are treated in the form of the Sudakov factors which
additionally suppress the end-point integration regions. The model
leads to the following hierarchy of the amplitudes
\begin{equation}
{\rm LL}:\;\;\;  {\cal M}_{0\,\nu,0\,\nu}^{V(g)} \propto
1\,;\qquad {\rm TL}:\;\;\;  {\cal M}_{0\,\nu,+\nu}^{V(g)} \propto
                                                  \frac{\sqrt{-t}}{Q};\qquad
{\rm TT}:\;\;\;{\cal M}_{+\nu,+\nu}^{V(g)} \propto \frac{\vk^2}{Q
M_V}.
 \end{equation}
The other transitions are small and  neglected in our analysis.
Here $M_V$ is the scale which appears in the higher order wave
function for transversely polarized meson and, respectively, in
the $TT$ amplitude  \cite{gk05}. The $M_V$ should be about the
 meson mass $m_V$.

The GPD is  complicated function which depends on three variables.
For the small momentum transfer  $t \sim 0$ we can use the double
distribution form proposed in \cite{mus99} \be H^{g}(\xb,\xi,t) =
\Big[\,\Theta(0\leq \xb\leq \xi)
    \int_{\frac{\xb-\xi}{1+\xi}}^{\frac{\xb+\xi}{1+\xi}}\, d\beta +
     \Theta(\xi\leq \xb\leq 1) \int_{\frac{\xb-\xi}{1-\xi}}
     ^{\frac{\xb+\xi}{1+\xi}}\, d\beta \,\Big]\,
 \frac{\beta}{\xi}\,f(\beta,\alpha=\frac{\xb-\beta}{\xi},t)
       \ee
with the simple factorizing ansatz for the double distributions
$f(\beta,\alpha,t)$

\be f(\beta,\alpha,t\simeq 0) = g(\beta)\,\frac{3}{4}\,
\frac{[(1-|\beta|)^2-\alpha^2]}{(1-|\beta|)^{3}}. \ee In this
model, we calculate GPD \cite{gk05} through  the gluon
distribution $g(\beta)$ and take the CTEQ5M results~\cite{CTEQ}
for it . Unfortunately, in the low $x$ region the gluon
distribution has an error which can exceed 15\%. This will  cause
large uncertainty in the calculated cross sections.

The  $t$- dependence of the amplitudes is important in analyses of
experimental data. For simplicity we parameterize it in the
exponential form $M_{ii}(t)=M_{ii}(0)\; e^{t\,B_{ii}/2}$ for $
LL,\; TT,\; TL$ transitions. Experimentally, only the slope of the
$\gamma^* p\to Vp$ cross section is measured whereas information
about individual slopes $B_{ii}$  is absent. We suppose   that  $
B_{LL} \sim B_{TL}$ but $B_{TT}$ can not be  equal to $B_{LL} $.
In the integrated cross section only the  following combination
$|M_{TT}|^2 \propto (\frac{f^V_T }{M_V})^2\frac{1}{B_{TT}}$
appears. Thus, by the corresponding choice of parameters we can
obtain the same cross sections for a different $B_{TT}$ slope. We
test two different
scenarios for the $\rho$ production which will be discussed here.\\
1. $B_{TT} \sim B_{LL}/2;\;M_V=m_V;\;f_{\rho
T}=.250\gev;\;a_{\rho T}=0.65\gev^{-1}$.\\
 2. $B_{TT} \sim B_{LL};\;M_V=m_V/2;\;f_{\rho
T}=.170\gev;\; a_{\rho T}=0.65\gev^{-1}$. \\
This will result in the same integrated over $t$ cross section.
Differences can be found only in observables with the interference
between different helicity amplitudes (SDME e.g.). For $LL$
transition we use $f_{\rho L}=.216\gev$,
  $a_{\rho L}=0.522\gev^{-1}$.

\begin{figure}[h!]
\begin{center}
\begin{tabular}{cc}
\mbox{\epsfig{figure=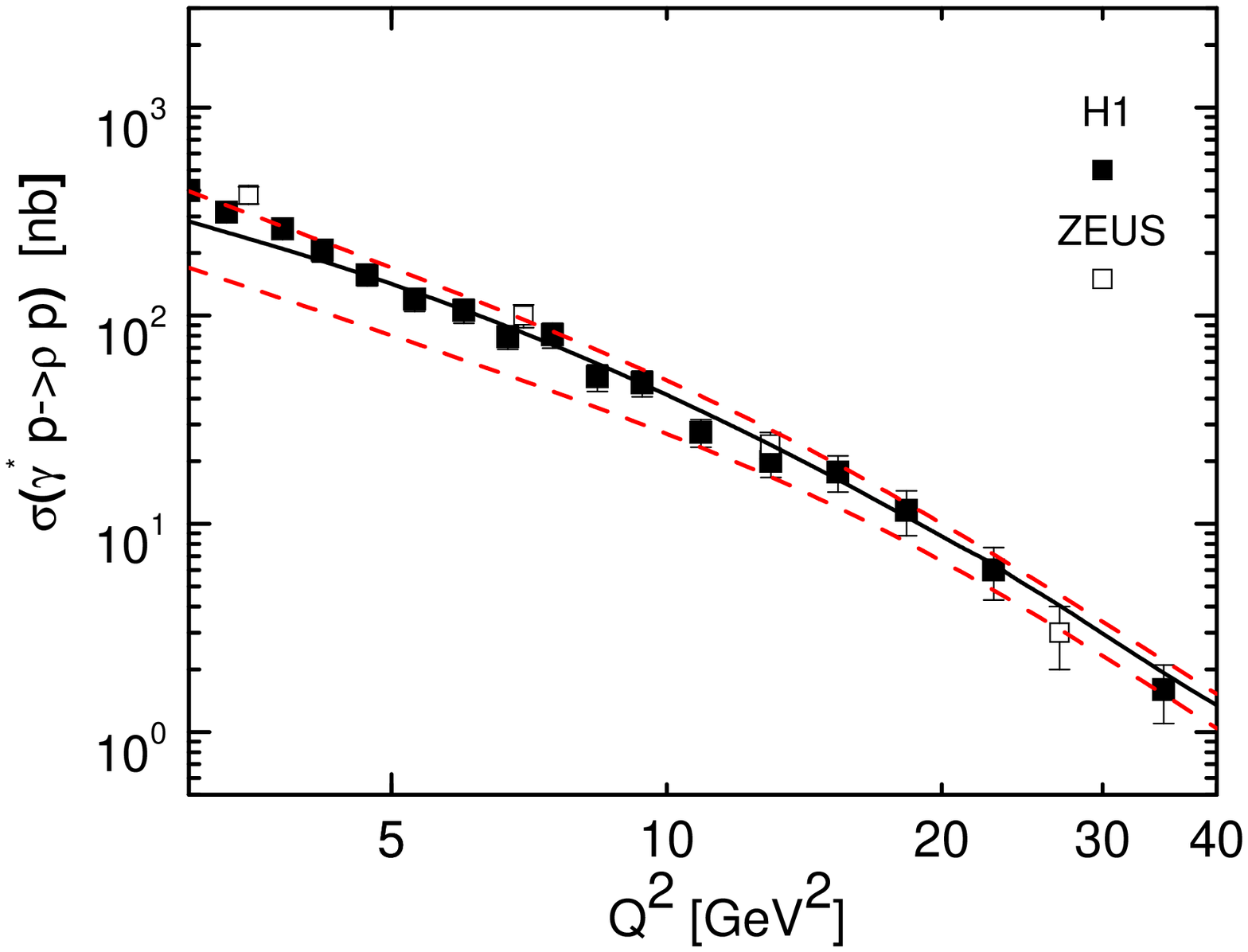,width=7.7cm,height=6cm}}&
\mbox{\epsfig{figure=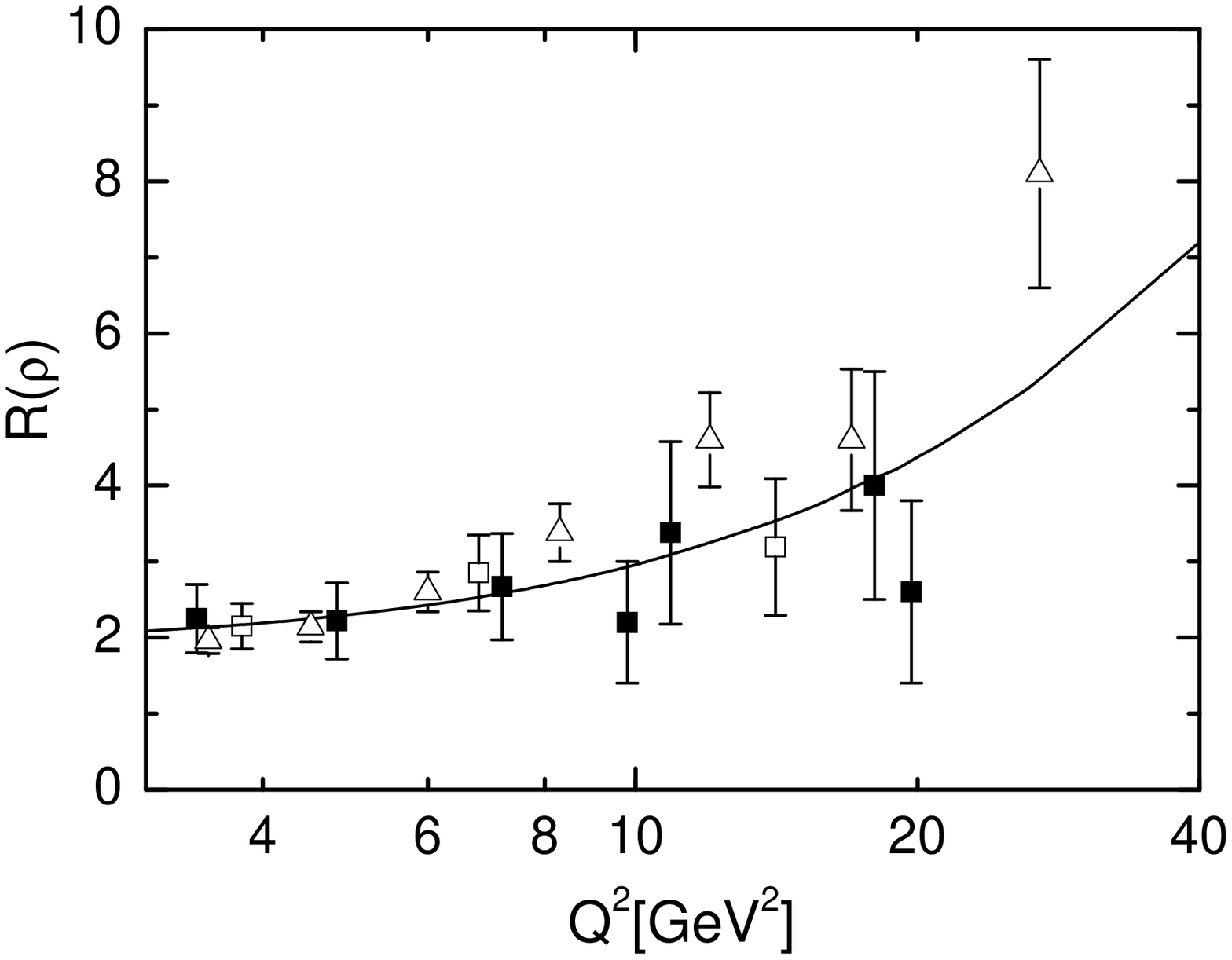,width=7.5cm,height=6.2cm}}\\
{\bf(a)}& {\bf(b)}
\end{tabular}
\end{center}
{\small{\bf Figure 2a.} The cross section for $\gamma^*\, p\to
\rho^0\, p$ vs. $Q^2$ for $W=75\gev$ (full line).  Dashed lines
represent errors in the cross sections
from uncertainty  at gluon distribution.}\\
{\small{\bf Figure 2b.} The ratio of longitudinal and transverse
cross sections
 for the $\rho$  production vs. $Q^2$ at
 $W =75\, \gev$ and $t=-0.15\,\gev^2$.
  The   line
 - our results. Data are  from H1 and ZEUS.}
\end{figure}

The cross section for the $\gamma^* p \to \rho p$
 production integrated over $t$  is shown
in Fig. 2a (full line). Good agreement with DESY experiments
\cite{h1,zeus} is to be observed. As mentioned before, the gluon
GPD has quite large errors. The dashed lines in Fig. 2a reflect
uncertainty in our results caused by the gluon distribution
$g(\beta)$ in (6). It can be seen that the obtained uncertainty
exceeds essentially experimental errors  in  the cross section at
small $Q^2$.

\begin{figure}[t!]
\begin{center}
\begin{tabular}{cccc}
\mbox{\epsfig{figure=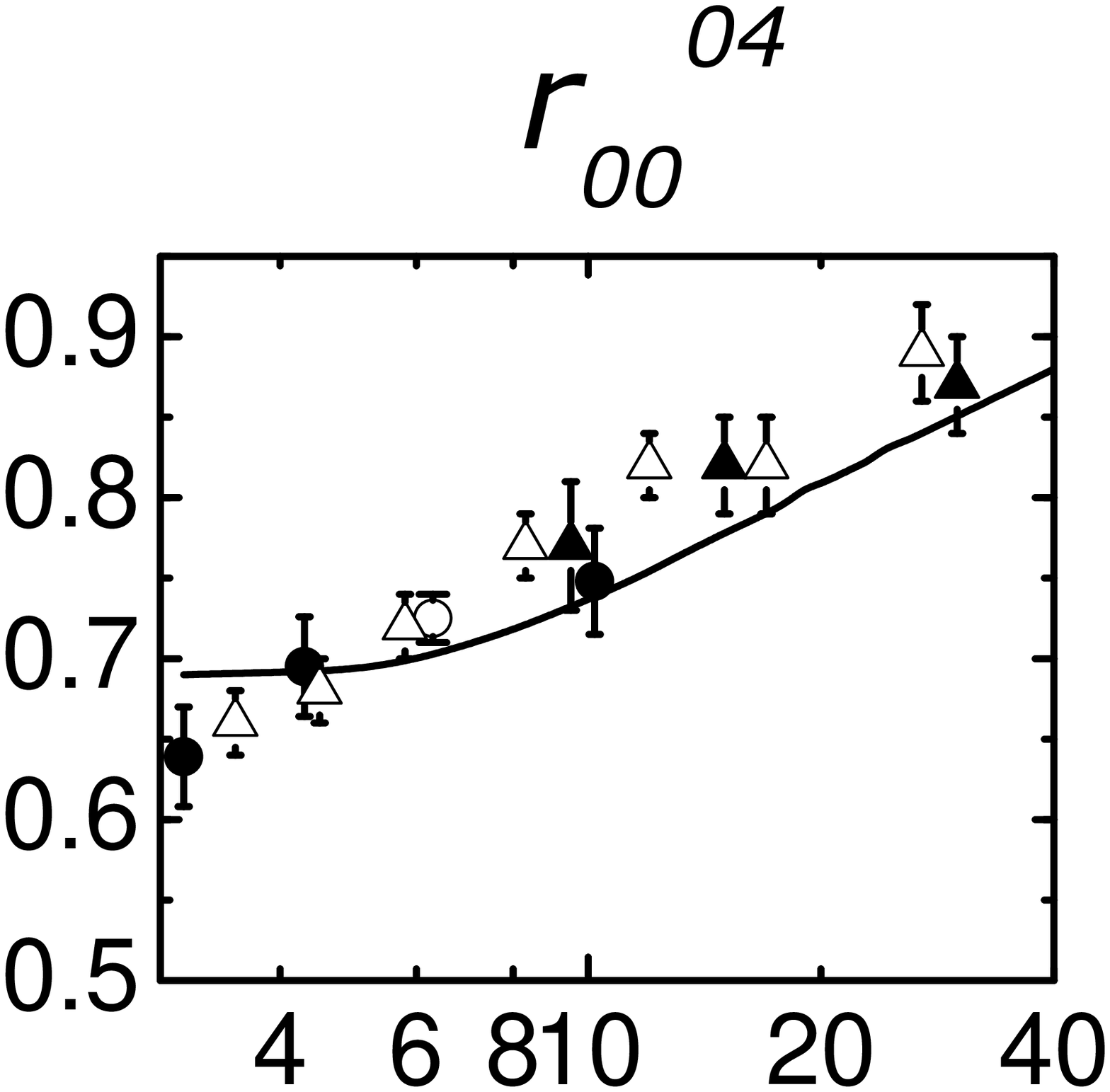,width=3.5cm,height=3.5cm}}&
\mbox{\epsfig{figure=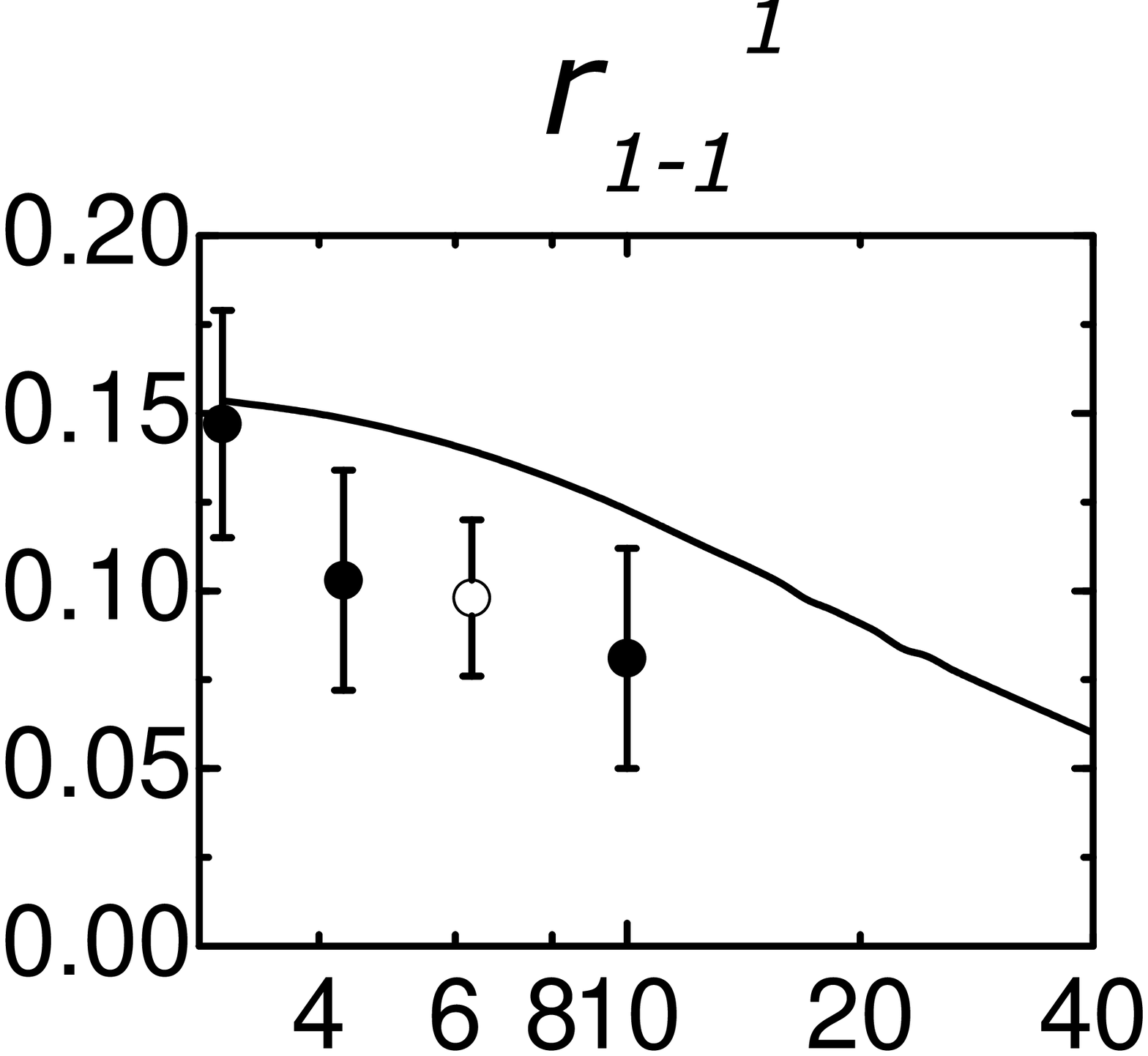,width=3.5cm,height=3.5cm}}&
\mbox{\epsfig{figure=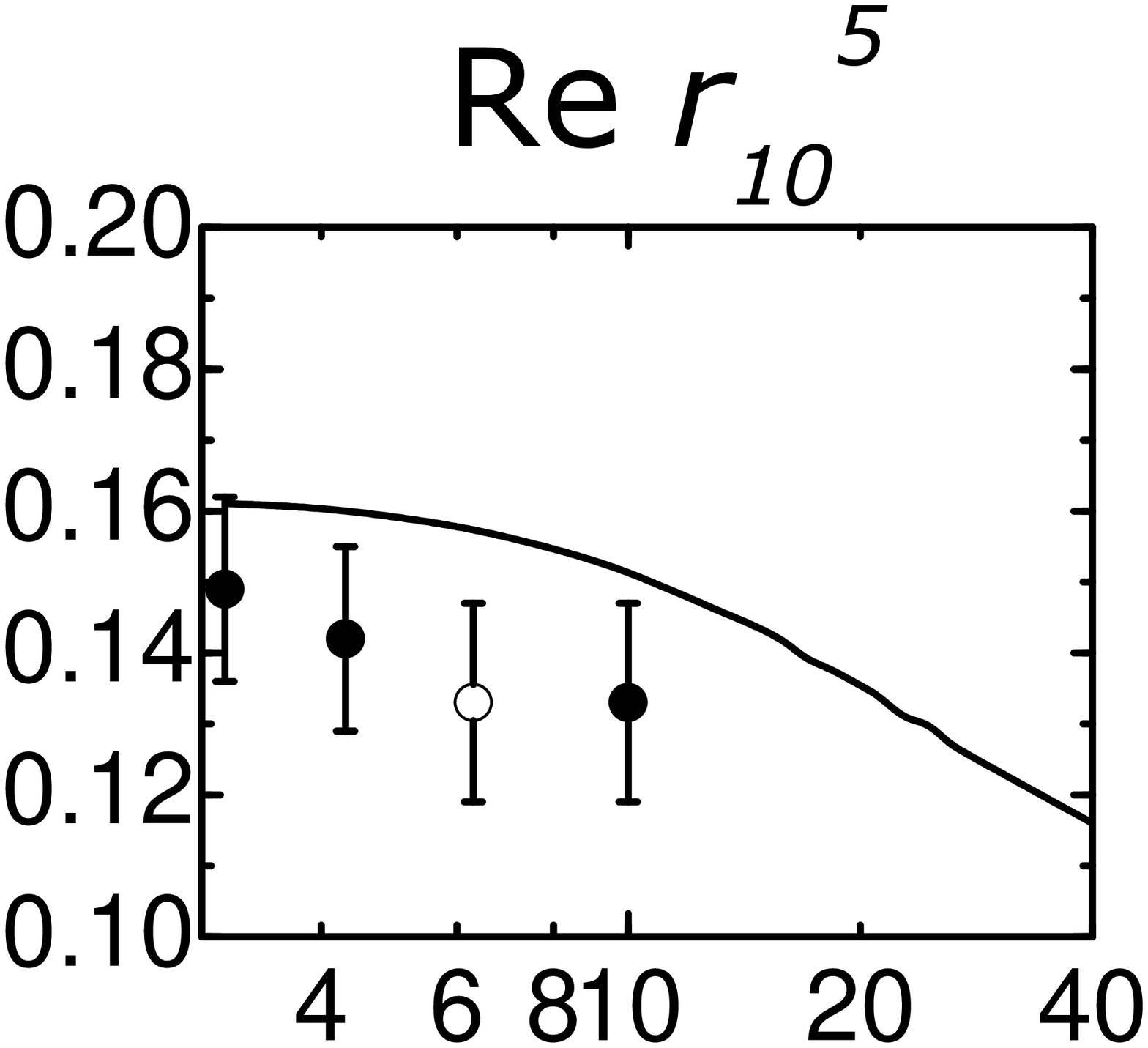,width=3.5cm,height=3.5cm}}&
\mbox{\epsfig{figure=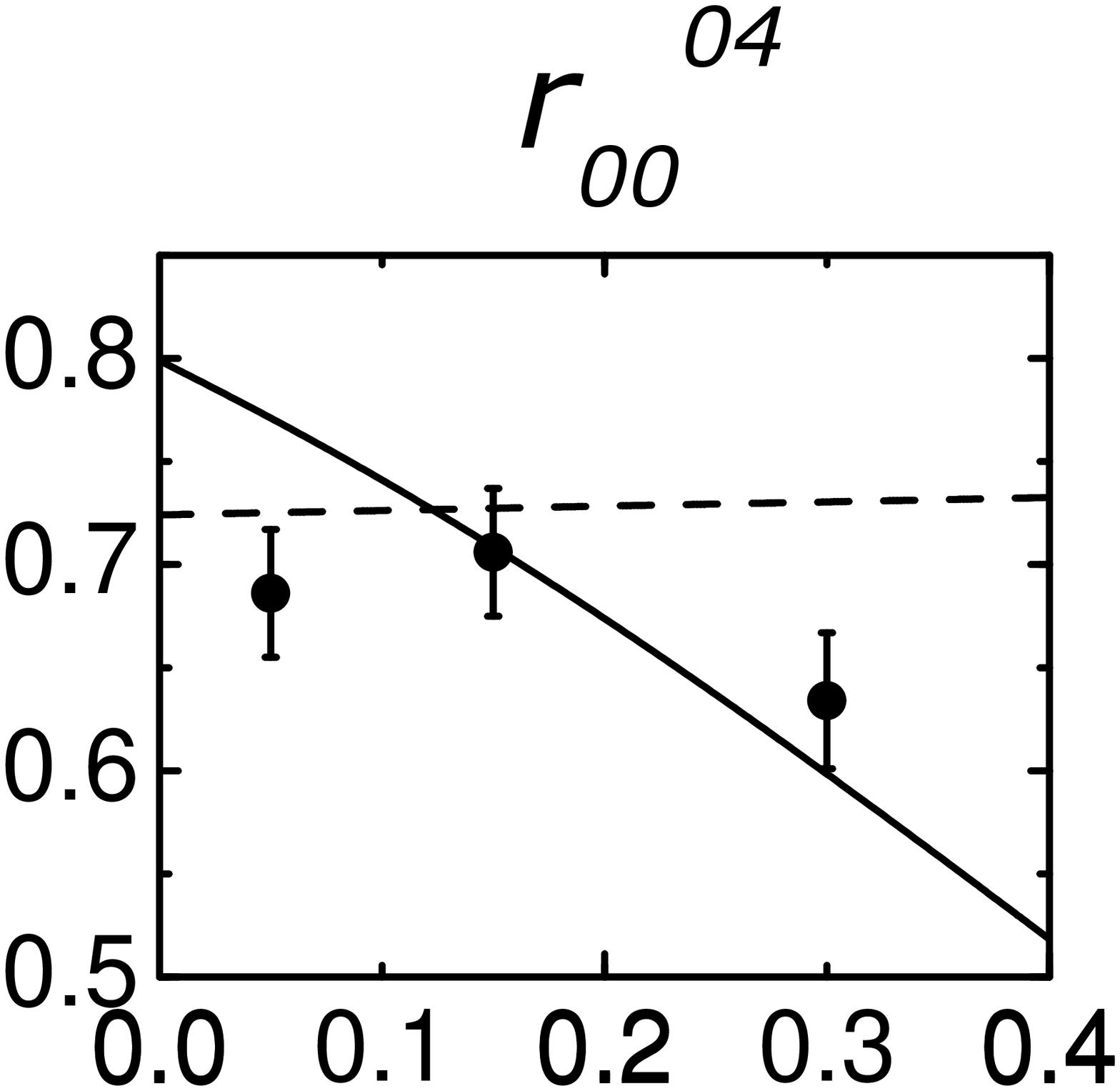,width=3.5cm,height=3.5cm}}\\
\end{tabular}
\phantom{aa}\vspace{-3mm}
\centerline{\hspace{4.3cm}$Q^2[\mbox{GeV}]^2$\hspace{6.1cm}-$t
[\mbox{GeV}]^2$}
\end{center}
{\small{\bf Figure 3} Three left figures: the $Q^2$ dependence of
SDME on the $\rho$ production at $t=-.15\gev^2$ and $W=75\gev$.
Right figure: The $t$- dependence of SDME on the $\rho$ production
at $Q^2=5\gev^2$ and $W=75\gev$.  Data are taken from H1 and ZEUS.
The solid (dashed) lines- our results for $B_{TT} =
B_{LL}/2\;(B_{LL}$).}
\end{figure}

The ratio of the cross section with longitudinal and transverse
photon polarization $R$ was calculated too. The model results for
the ratio $R$ of the $\rho$ production is shown in Fig. 2b and in
consistent with H1 and ZEUS  experiments  \cite{h1,zeus}.

In Fig.3, we present our results for   SDME at DESY energy range.
The three left figures show the $Q^2$ dependencies of SDME.
Description of experimental data is reasonable. In the right
figure of Fig. 3 we show the $t$ -dependence of $r_{00}^{04}$ SDME
for different scenarios 1,2. They give quite different predictions
for the momentum dependence of $r_{00}^{04}$  but both the  models
agree with the known H1 experimental data \cite{h1} at small
momentum transfer.

\begin{wrapfigure}[16]{hr}{5.5cm}
\begin{center}
\mbox{\epsfig{figure=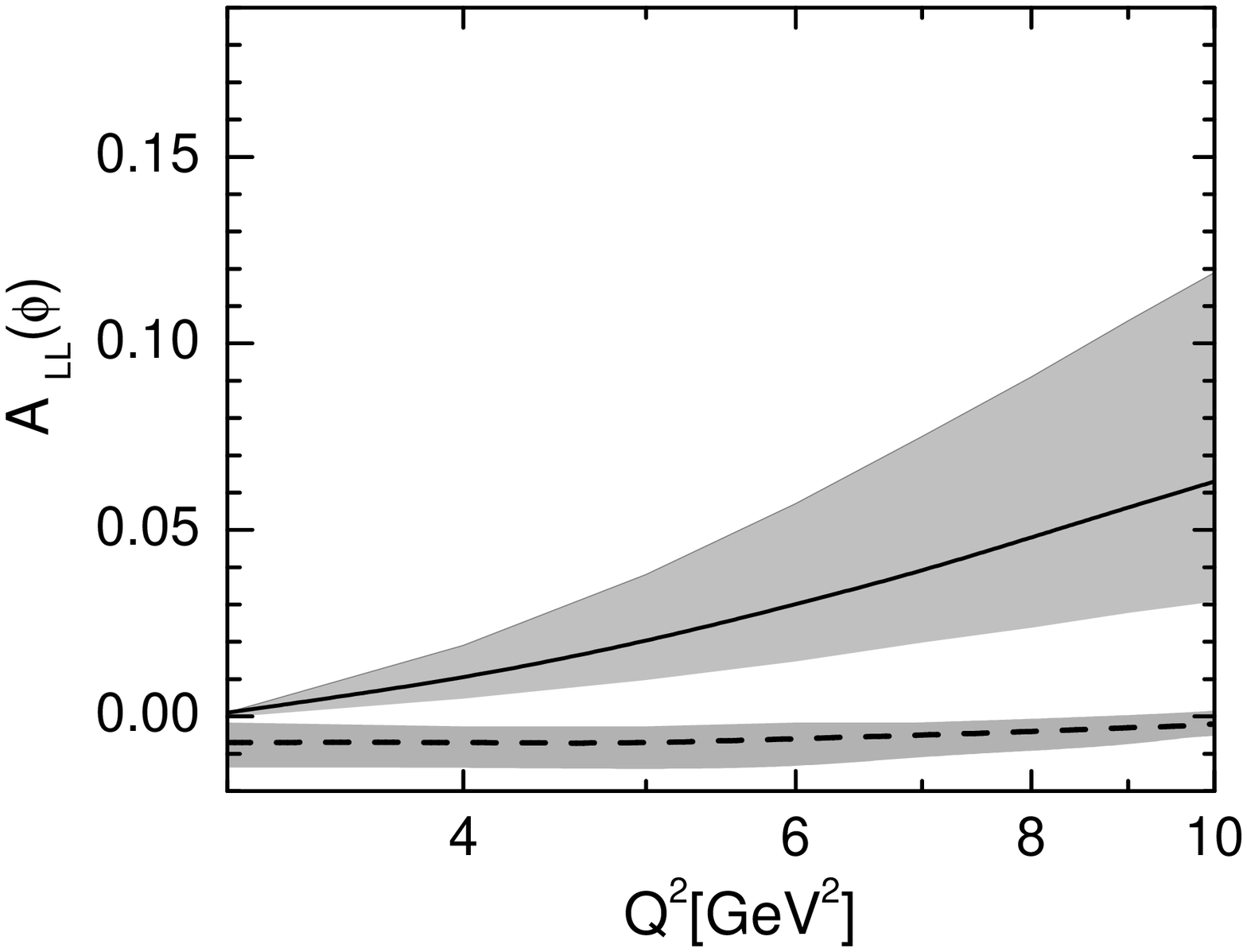,width=5.2cm,height=4cm}}
\end{center}
{\small{\bf Figure 4.} $A_{LL}$ for the $\phi$ production at
$W=5\,\gev$ (solid line) and
  $W=10\,\gev$ (dashed line); $y\simeq 0.6$. The shaded bands reflect
  the uncertainties in $A_{LL}$ due to the  error in
  $\widetilde{H}^g \propto \Delta g$  }
  \label{yourname_fig4}
\end{wrapfigure}
The $A_{LL}$ asymmetry for a longitudinally polarized beam and
target is sensitive to the polarized GPD.  Really, the leading
term in $A_{LL}$ asymmetry integrated over the azimuthal angle is
determined through the interference between the $H^g$ and the
$\widetilde{H}^g$ distributions
$$A_{LL}[ep\to epV] = 2 \sqrt{1-\veps^2}\,
                \frac{{\rm Re}\;\Big[ {\cal M}^H_{++,++}\,
                 {\cal M}^{\widetilde{H}*}_{++,++}\Big]}
               {\veps |{\cal M}^H_{0+,0+}|^2 + |{\cal
               M}^H_{++,++}|^2}\,.$$
We expect a small  value for $A_{LL}$ asymmetry at high energies
because it is of the order of the ratio $\langle
\widetilde{H}^g\rangle/\langle H^g \rangle$ which is small  at low
$x$.

At COMPASS(SMC) energies $W=10 \mbox{GeV}$($15 \mbox{GeV}$) the
$A_{LL}$ asymmetry of the $\phi$ production is expected to be
small (Fig.4). At HERMES energies $W=5 \mbox{GeV}$ the major
contribution to the amplitudes comes  from the region $0.1 \le \xb
\le 0.2$ where $\Delta g/g$ is not small, which leads to a large
value of $A_{LL}$. The asymmetry  is shown in Fig.4 together with
the uncertainties in our predictions due to the error in the
polarized gluon distribution. Note that for the $\rho$ production
the polarized quark GPD should be essential in the $A_{LL}$
asymmetry.

{\bf In summary}: Light vector meson electroproduction at small
$x$ was analyzed here within the  GPD approach. The amplitudes
were calculated  using the MPA with the wave function  dependent
on the transverse quark momentum. By including  the higher order
$k_\perp^2/Q^2$  effects in the denominators of hard subprocess
(\ref{hsaml}) we regularize the end-point singularities in the
amplitudes with transversally polarized photons.   It was pointed
out that the diffraction peak slopes of the $TT$ and $TL$
amplitudes are not well defined. However, within scenarios 1 and 2
we can get the same accurate description of the $Q^2$ dependence
of the cross section, $R$ ratio and   SDME for the $\rho$ meson
production. Our results for the $\phi$ production can be found in
\cite{gk05}. Some comparison with the two-gluon exchange model
\cite{bro94} known as $\ln(1/x)$ approximation can be found in
\cite{gk05} too.

The knowledge of the $B_{TT}$ and $B_{TL}$ slopes is essential in
analyses of the momentum transfer dependence of the SDME. We found
 reasonable results for SDME in the GPD model for both the scenarios.
This means that now we have two possibilities for the $t$-
dependence of the scattering amplitudes with similar and different
$B_{LL}$ and $B_{TT}$ slopes which agree with existing
experimental data. Unfortunately, the data on spin observables
have  large experimental errors. This does not permit one to
determine which model is relevant to experiment. To clarify
situation,  additional theoretical study of the momentum
dependence of the $LL$, $TT$ and $TL$ transition amplitudes is
needed. An experimental investigation to reduce errors in SDME is
extremely important.

Thus, we can conclude that the vector meson photoproduction at
small $x$ is a good tool to probe the gluon  GPD. Study  of the
$t$ dependence of SDME can give  important information on the
structure of different
helicity amplitudes in the vector meson production.\\

This work is supported  in part by the Russian Foundation for
Basic Research, Grant 03-02-16816  and by the Heisenberg-Landau
program.

\end{document}